\documentclass[journal]{IEEEtran}
\usepackage{amsmath}
\usepackage{amssymb}
\usepackage{subfigure}
\usepackage{graphicx}
\usepackage{multirow}
\usepackage{booktabs}
\usepackage{makecell}
\usepackage{cite}
\usepackage{array}
\usepackage{caption}
\usepackage{supertabular}
\usepackage{color}
\usepackage{stfloats}
\usepackage{algorithm}
\usepackage{algpseudocodex}
\def\BibTeX{{\rm B\kern-.05em{\sc i\kern-.025em b}\kern-.08em
    T\kern-.1667em\lower.7ex\hbox{E}\kern-.125emX}}
\algrenewcommand\algorithmicrequire{\textbf{Initialize:}}
\algrenewcommand\algorithmicensure{\textbf{Output:}}
\algnewcommand{\algorithmiciteration}{\textbf{Iteration:}}
\makeatletter
\renewcommand{\floatc@ruled}[2]{{\@fs@cfont #1:} #2\par}
\algnewcommand\Iteration{%
    \algpx@endCodeCommand%
    \ifnumcomp{0}{<}{\FSSize{algpx@startNewCodeBoxQueue}}{\setbool{algpx@setNorth}{true}}{}%
    \algpx@drawInItem{\algorithmiciteration}%
}
\makeatother

\ifCLASSINFOpdf
\else
\fi

\begin{document}
\bstctlcite{IEEEexample:BSTcontrol}
\bibliographystyle{IEEEtran}
\captionsetup[figure]{labelfont={normalfont},labelformat={default},labelsep=period,name={Fig.}}
\captionsetup{font={small}}
%
\title{\LARGE Analog Beamforming for In-Band Full-Duplex Phased Arrays with Quantized Phase Shifters under a Per-Antenna Received Power Constraint}
%
%
%

\author{Ao~Liu, 
Ian~P.~Roberts,Taneli~Riihonen, and Weixing~Sheng}%
\maketitle

\begin{abstract}
This letter develops a novel transmit beamforming (BF) design for canceling self-interference (SI) in analog in-band full-duplex phased arrays. 
Our design maximizes transmit BF gain in a desired direction while simultaneously reducing SI power to below a specified threshold on per-antenna basis to avoid saturating receive-chain components, such as LNAs.
Core to our approach is that it accounts for real-world phase shifters used in analog phased array systems, whose limited resolution imposes non-convex constraints on BF design.
We overcome this by transforming these non-convex constraints into convex polygon constraints, which we then solve through semidefinite relaxation and a rank refinement procedure.
Numerical results show that our proposed BF scheme reliably cancels SI to the target power threshold at each receive antenna while sacrificing little in transmit BF gain, even with modest phase shifter resolution.

\end{abstract}


%
\IEEEpeerreviewmaketitle

\section{Introduction}
\label{Introduction}
\IEEEPARstart{P}{hased} array systems with in-band full-duplex (IBFD) \cite{1} capability can transmit (Tx) and receive (Rx) simultaneously at the same carrier frequency. 
This can increase spectral efficiency in wireless communication systems \cite{4,5} and can unlock functionalities such as enhanced continuous-wave sensing \cite{2,3}, integrated sensing and communications \cite{6,7}, and spectrum sensing.

It is well known that enabling an IBFD phased array system requires successful cancellation of the self-interference (SI) that its transmitter inflicts onto its own co-located receiver \cite{8,carlos}.
Regardless of the particular SI cancellation (SIC) technique (e.g., analog, digital), it is essential that SI be kept sufficiently low at each Rx antenna to avoid overwhelming front-end components, such as low-noise amplifiers (LNAs), which saturate beyond a certain input power level \cite{lna,roberts_bflrdr}.
This introduces the notion of a per-antenna SI power constraint, which we address in this work with the understanding that other SIC mechanisms can be applied atop our proposed technique to thoroughly eliminate SI.

It is often assumed that a radio frequency (RF) filter can be used to ensure SI is kept sufficiently low at each Rx antenna, but the cost, complexity, and size of such solutions can make them impractical when scaled to phased arrays with dozens of antennas.
With this in mind, we instead propose harnessing Tx beamforming (BF) to not only deliver high BF gain but also mitigate SI on a per-antenna basis, leveraging ample spatial degrees-of-freedom afforded by dense phased arrays \cite{9}.

Rather than performing BF digitally, most real-world phased array systems, such as those in 5G \cite{10,5g}, employ an \textit{analog} BF architecture, where RF phase shifter components are used to physically realize desired BF weights. 
Such analog BF architectures have practical advantages, but they are hindered by the limited phase resolution of phase shifters and, often, the lack of amplitude control.
While a number of techniques have been developed specifically for SIC in analog BF systems, many ignore these practical considerations.

For instance, \cite{13,pafd,15} assume continuous, infinite-resolution phase shifters, capable of realizing any desired phase shift. 
Consequently, these designs are not faced with optimizing their BF weights over the non-convex set of phase shifter settings.
The works of \cite{11,12}, on the other hand, account for limited-resolution phase shifters but also assume amplitude control in their designs, which is not always available in practice.
Works such as \cite{17,18,19,20} do in fact account for both discrete phase shifters and a lack of amplitude control, but fall short in other respects.
For example, the design of \cite{17} uses exhaustive search to configure discrete phase shifter settings and that of \cite{18} uses a genetic algorithm (GA).
While these approaches prove to be effective in BF design with limited-resolution phase shifters, their complexity is extremely high, even for a modest number of antennas. 
In \cite{19,20}, sequential algorithms are used, which---when applied to our problem---will effectively relax the non-convexity of limited-resolution phase shifters and then project the solution onto the set of discrete phase shifter settings.
While this may be suitable in many BF applications, naively projecting a desired solution can be detrimental in one's attempt to cancel SI and enable IBFD operation, since only a modest degree of residual SI can make IBFD infeasible.

Altogether, this motivates the work presented herein, which proposes an analog Tx BF design which mitigates SI at each Rx antenna while explicitly accounting for the limited phase control and lack of amplitude control in practical analog arrays.
Our design maximizes Tx BF gain in a desired direction while ensuring the SI power inflicted onto each Rx antenna is below some specified threshold. 
In doing so, we account for quantized phase shifters by transforming the non-convex phase constraint to a convex polygon constraint. 
We then use semidefinite relaxation (SDR) along with a rank refinement and phase rotation procedure to reliably solve for the near-optimal Tx BF weights.
Numerical results validate the effectiveness of the proposed method and its superiority over existing methods, minimally trading off Tx BF gain for SI mitigation per-antenna to preserve linearity of the Rx chain.


\begin{figure}[tp]
	\centering
		{\includegraphics [width=0.38\textwidth]{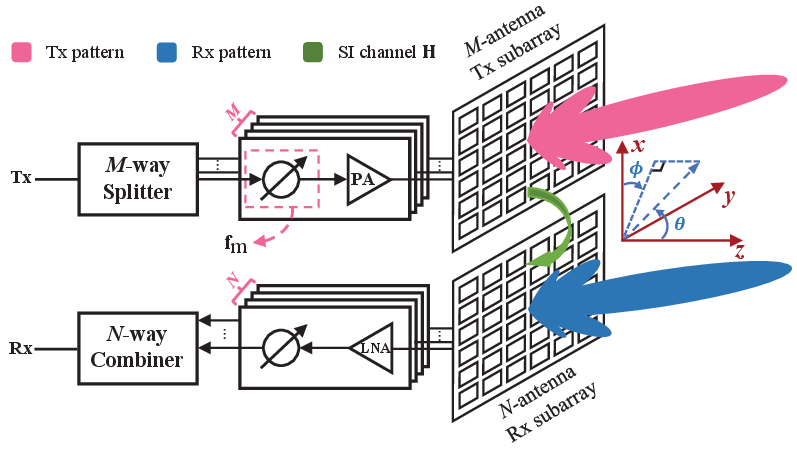}}
    \caption{The IBFD phased array architecture considered in this work, employing an analog BF network of limited-resolution phase shifters.}
	\label{arraymodel}
\end{figure}

\section{System Model}\label{System Model} 
Consider the IBFD analog phased array illustrated in Fig.~\ref{arraymodel}, whose aperture is partitioned into an $M$-antenna Tx subarray and an adjacent $N$-antenna Rx subarray. 
Such a partitioning has been found to be more favorable for IBFD operation than other partitionings \cite{21,22}.
Each subarray is equipped with a single RF chain and is electronically steered via a BF network of digitally controlled phase shifters. 

The elevation and azimuth angles relative to the array boresight are denoted as $\theta$ and $\phi$ in Fig.~\ref{arraymodel}, respectively, and let $\mathbf{a}(\theta,\phi)\in\mathbb{C}^{M\times1}$ denote the array response vector in the direction $(\theta,\phi)$. 
Let $\mathbf{f}\in\mathbb{C}^{M\times1}$ be the Tx BF weights, and we denote by $\mathbf{f}_m$ the $m$-th element of $\mathbf{f}$. 
The total transmitted power is $P_\mathrm{t} = {\|\mathbf{f}\|}^2_2$, where $\Vert \mathbf{x}\Vert_2$ denotes the $\ell_2$-norm of $\mathbf{x}$. 
To physically realize a BF vector $\mathbf{f}$ with digitally controlled phase shifters, it must have entries of the form 
\begin{equation}
\mathbf{f}_m = \sqrt{P_{\mathrm{t}}/M} \cdot \exp(\mathrm{j} \cdot \varphi_m),
\end{equation}
where the phase $\varphi_m \in \mathcal{V}$ is an element of the discrete set 
\begin{equation}\label{eq2}
\mathcal{V}=\left\{v_k| v_k=\frac{2\pi\cdot(k-1)}{K},\ k=1,2,\ldots,K\right\},
\end{equation}
where $K = 2^b$ denotes the number of phase shifter settings and $b$ is the phase shifter resolution in bits. 
Lacking amplitude control, the total Tx power is split evenly across all $M$ antennas and thus $|\mathbf{f}_m|^2 = {P_\mathrm{t}}/{M},\ \forall m=1,2,\ldots,M$.
Shown in Fig.~\ref{arraymodel}, the SI channel matrix $\mathbf{H}\in\mathbb{C}^{N\times M}$ captures the coupling between each pair of Tx and Rx antennas.

\section{Problem Formulation and Proposed Method}\label{Design Approach}
In this letter, we aim to optimize the Tx BF weights $\mathbf{f}$ to suppress the SI power at each of the $N$ Rx antennas while maintaining high BF gain in a desired direction $(\theta,\phi)$.
Given the established system model, we first quantify our design criteria and formulate the corresponding expressions. Subsequently, the detailed design problem is formulated by assembling these expressions.

The BF gain towards some direction of interest $(\theta,\phi)$ afforded by $\mathbf{f}$ can be expressed as 
\begin{equation}\label{eq5}
G(\theta,\phi) = \frac{|\mathbf{a}^*(\theta,\phi)\mathbf{f}|^2}{\displaystyle\mathbf{f}^{*}\mathbf{f}},
\end{equation}
which is maximized by conjugate BF (CBF) when $\mathbf{f} = \sqrt{P_{\mathrm{t}}/M} \cdot \mathbf{a}(\theta,\phi)$; here, $\mathbf{x}^*$ denotes the conjugate transpose of $\mathbf{x}$. 
Meanwhile, the SI power inflicted onto the $n$-th Rx antenna by $\mathbf{f}$ can be expressed as 
\begin{equation}\label{eq4}
P^{\mathrm{SI}}_n = \mathbb{E}\left[ \|\mathbf{H}_{n}\mathbf{f}\|^2_2\right] = \mathbf{f}^*\mathbf{C}_n\mathbf{f},
\end{equation}
where $\mathbf{H}_n$ is the $n$-th row of $\mathbf{H}$ and $\mathbf{C}_n = \mathbf{H}_n^*\mathbf{H}_n \in \mathbb{C}^{M \times M}$. 

As outlined before, we aim to design the Tx BF weights $\mathbf{f}$ which mitigate the per-antenna SI power $P^{\mathrm{SI}}_n \ \forall n$ while delivering high BF gain $G(\theta,\phi)$. 
In this pursuit, assemble a Tx BF design problem as
\begin{subequations}\label{eq6}
\begin{align}
\mathcal{P}_1:\ \max_{\mathbf{f}}  & \ G(\theta,\phi)  \tag{\ref{eq6}}\\
\mathrm{s.t.} &\ |\mathbf{f}_m|^2 =\frac{P_\mathrm{t}}{M},\ \forall m=1,2,\ldots,M \label{eq6:a}\\[0.5em] 
& \ P^{\mathrm{SI}}_n \leq P_\mathrm{max} ,\ \forall n=1,2,\ldots,N\label{eq6:b}\\[0.5em]
& \ \arg(\mathbf{f}_m) \in \mathcal{V} ,\ \forall m = 1,2,\ldots,M\label{eq6:c}
\end{align}
\end{subequations}
where $P_\mathrm{max}$ is the maximum SI power tolerated at  each antenna and $\mathrm{arg}(x)$ denotes the phase of $x$. 
Maximizing $G(\theta,\phi)$ in the objective of \eqref{eq6} is non-convex but can be easily reformulated as the quadratic function minimization 
\begin{equation}\label{eq8}
\min_{\mathbf{f}}   \ \mathbf{f}^*\mathbf{A}\mathbf{f}
\end{equation}
where $\mathbf{A} = -\mathbf{a}\mathbf{a}^* \in \mathbb{C}^{M \times M}$ and $\mathbf{f}^*\mathbf{f} = P_\mathrm{t}$. 
Still, problem $\mathcal{P}_1$ remains NP-hard, as minimizing a quadratic function with a negative semidefinite matrix $\mathbf{A}$ is non-convex. 
In light of this, we transform $\mathcal{P}_1$ into a semidefinite programming (SDP) problem with positive semidefinite (PSD) matrix $\mathbf{F} = \mathbf{f}\mathbf{f}^*$ as 
\begin{subequations}\label{eq9}
\begin{align}
\mathcal{P}_2:\ \min_{\mathbf{F} = \mathbf{f}\mathbf{f}^*}  & \ \mathrm{tr}(\mathbf{A}\mathbf{F})  \tag{\ref{eq9}}\\
\mathrm{s.t.} &\ \mathbf{F}_{m,m} = \frac{P_\mathrm{t}}{M},\ \forall m=1,2,\ldots,M \label{eq9:a}\\[0.5em] 
& \ \mathrm{tr}(\mathbf{C}_n\mathbf{F})\leq P_\mathrm{max} ,\ \forall n=1,2,\ldots,N\label{eq9:b}\\[0.5em]
& \ \arg(\mathbf{f}_m) \in \mathcal{V} ,\ \forall m = 1,2,\ldots,M\label{eq9:c}\\[0.5em] 
& \ \mathbf{F} \succeq \mathbf{0}, \ \mathrm{rank}(\mathbf{F}) = 1 \label{eq9:d}
\end{align}
\end{subequations}
where $\mathrm{tr}(\cdot)$ denotes the trace operation and $\mathbf{F} \succeq \mathbf{0}$ and $\mathrm{rank}(\mathbf{F}) = 1 $ are imposed to guarantee that the factorization $\mathbf{F} = \mathbf{f}\mathbf{f}^*$ exists. 
The SDR method can be directly implemented here to relax the non-convex rank-$1$ constraint, which leaves (\ref{eq9:c}) as the only remaining non-convex constraint. 

Next, we relax the discrete phase constraint (\ref{eq9:c}) into a convex constraint by recognizing that the $K$ phase shifts in $\mathcal{V}$ are distributed uniformly around the unit circle with radius $\sqrt{{P_\mathrm{t}}/M}$.
The convex hull of these phase shifts is a regular polygon with $K$ vertices, with each vertex a point in $\mathcal{V}$ \cite{23}. 
This can be formalized as constraining $\mathbf{f}_m \in \mathcal{F}_{\mathcal{V}} \ \forall m$ where
\begin{multline} \label{eq10}
\mathcal{F}_{\mathcal{V}}=
\Big\{\mathbf{f}_m | \operatorname{Re}\left(t_k^* \mathbf{f}_m\right) \leq \sqrt{\frac{P_\mathrm{t}}{M}} \cos\left(\frac{\pi}{K}\right), t_k=\mathrm{e}^{\mathrm{j} \frac{\left(v_k+v_{k+1}\right)}{2}}, \\  k=1,2, \ldots, K \Big\}
\end{multline}
is the convex set containing the intersection of $K$ halfspaces defined by $K$ sides of the regular polygon, with $t_k$ the outer normal of the line between phase $v_k$ and $v_{k+1}$. 
Notice that this regular polygon is closed, where $v_1 = 0$ and $v_{K+1} = v_1$. 
With (\ref{eq10}), problem $\mathcal{P}_2$ can be transformed as 
\begin{subequations}\label{eq11}
\begin{align}
\mathcal{P}_3:\ \min_{\mathbf{F}}  & \ \mathrm{tr}(\mathbf{A}\mathbf{F})  \tag{\ref{eq11}}\\
\mathrm{s.t.} &\ \mathbf{F}_{m,m} = \frac{P_\mathrm{t}}{M},\ \forall m=1,2,\ldots,M \label{eq10:a}\\[0.5em] 
& \ \mathrm{tr}(\mathbf{C}_n\mathbf{F})\leq  P_\mathrm{max} ,\ \forall n=1,2,\ldots,N\label{eq11:b}\\[0.5em]
& \ \mathbf{F}_{p,q} \in \mathcal{F}_{\mathcal{V}} ,\ p > q \geq 1   \label{eq11:c}\\[0.5em] 
& \ \mathbf{F} \succeq \mathbf{0} \label{eq11:d}
\end{align}
\end{subequations}
where the constraint $\mathbf{f}_{m} \in \mathcal{F}_{\mathcal{V}}$ is modified as (\ref{eq11:c}) to accommodate the SDR formulation by constraining the entries of the strict lower triangle of $\mathbf{F}$ to be in $\mathcal{F}_\mathcal{V}$.

Problem $\mathcal{P}_3$ is convex and can be solved by efficient, off-the-shelf solvers like CVX \cite{24} for an optimal PSD solution $\mathbf{F}^{\star}$ \cite{25}. 

With that being said, recognize that solving $\mathcal{P}_3$ will not necessarily lead to a solution $\mathbf{F}^\star$ which is rank-$1$ and thus cannot always be factorized as $\mathbf{F}^\star = \mathbf{f}\mathbf{f}^*$ to obtain $\mathbf{f}$. 
To overcome this, we employ a rank refinement method \cite{26} when solving problem $\mathcal{P}_3$ to reliably produce a PSD solution $\mathbf{F}^\star$ which is (approximately) rank-$1$.
Then, we take the singular value decomposition (SVD) of $\mathbf{F}^\star$ and select the left singular vector $\mathbf{f}^\star$ corresponding to the largest singular value of $\mathbf{F}^{\star}$; with $\mathbf{F}^{\star}$ approximately rank-$1$, $\mathbf{F}^{\star} \approx \mathbf{f}^\star{\mathbf{f}^\star}^*$ holds. 

Finally, since $\mathbf{f}^\star$ may not exactly satisfy the discrete phase constraint of \eqref{eq9:c} and thus may not be physically realizable, we find the vector $\widehat{\mathbf{f}}^\star$ whose phase is within $\mathcal{V}$ and nearest to the phase of $\mathbf{f}^\star$.
Before performing this non-convex projection, however, it is important to recognize that, for any solution $\mathbf{f}^\star$ which factorizes $\mathbf{F}$, an arbitrarily rotated version $\mathrm{e}^{\mathrm{j}\beta} \cdot \mathbf{f}^\star$ also factorizes $\mathbf{F}$ for any $\beta$.
Thus, we can find the most suitable rotation $\beta$ \textit{before} projecting $\mathbf{f}^\star$ to be physically realizable, which, in our case, is that which minimizes the maximum SI power across Rx antennas. 
Formally, this rotation and projection process can be expressed as solving
\begin{equation}\label{eq12}
\mathcal{P}_4: \ \min_{\beta}   \ \lVert \mathbf{H}\, \widehat{\mathbf{f}}^\star\rVert_{\infty}
\end{equation}
where $\lVert \mathbf{x} \rVert_{\infty}$ is the max absolute entry of $\mathbf{x}$ and the $m$-th entry of $\widehat{\mathbf{f}}^\star$ is $\widehat{\mathbf{f}}^\star_m = \sqrt{P_\mathrm{t}/M} \cdot \exp\left(\mathrm{j}\widehat{\varphi}^\star_m\right)$, with its phase found via
\begin{equation}\label{eq13}
\hat{\varphi}_m^\star=\arg \min _{v \in \mathcal{V}} \ \left|\arg \left(\mathbf{f}^\star_m\right) \cdot \beta -v\right|, \ m=1,2,\ldots,M.
\end{equation}
Problem $\mathcal{P}_4$ can be easily solved by exhaustive search with trivial complexity, yielding the physically realizable Tx BF weights $\widehat{\mathbf{f}}^{\star}$ which maximize BF gain while preventing SI from saturating components at each Rx antenna.
A summary of our entire optimization process is detailed in Algorithm~\ref{algo: 1}.

\section{Performance Analysis}\label{Simulation Results}

To assess our proposed design we conducted numerical simulation using parameters and models aligned with real-world systems. 
We consider a $28$ GHz 12$\times$6 antenna array with half-wavelength spacing, akin to one that may be used in 5G cellular systems, which we assume is partitioned into separate 6$\times$6 Tx and Rx subarrays stacked side-by-side with the Tx subarray on right of the Rx, when looking outward (along the $z$-axis) from the array. 
To simulate a realistic SI channel matrix $\mathbf{H}$, we model the array as a U-slot patch array in Ansys HFSS and use computational electromagnetics to realize the coupling between antenna elements. 
We use a Tx power of $P_\mathrm{t} = 30$~dBm and will vary the per-antena power threshold $P_\mathrm{max}$ across $ \left[-20, -8\right]$~dBm. 
We consider phase shifters with $b \in [3,6]$ bits of resolution, in line with practical phase shifter components.
We will steer the Tx beam across $\theta \in [0^\circ,40^\circ]$ and $\phi \in [0^\circ,360^\circ] $ to broadly assess our design's ability to suppress SI while delivering high BF gain.

As the most suitable benchmark, we compare our proposed design against the sequential method in \cite{19,20}. 
While \cite{19,20} cancels SI to enable IBFD operation, it does not do so on a per-antenna basis and, in the context of our problem, would result in a simple projection of a fully-digital BF design onto the set of discrete phase shifts.
We also compare against conventional CBF (denoted as Conv.) and fully-digital BF \cite{9} (denoted as Dig.), which serve as a useful upper bound on BF gain and a lower bound on SI power, respectively. 

\begin{algorithm}[t]
 \caption{Proposed Tx BF design with limited-resolution phase shifters under a per-antenna SI power constraint.}
 \label{algo: 1}
 \begin{algorithmic}[1]
 \renewcommand{\algorithmicrequire}{\textbf{Input:}}
 \Require desired steering direction $(\theta,\phi)$, SI channel matrix $\mathbf{H}$, SI power threshold $P_\mathrm{max}$, phase shifter resolution $b$.
  \State Solve problem $\mathcal{P}_3$ in (\ref{eq11}) for the optimal PSD solution $\mathbf{F}^{\star}$, using rank refinement \cite[Algorithm~1]{26} to ensure it is (approximately) rank-$1$.
\State Perform SVD on $\mathbf{F}^\star$ to extract the left singular vector $\mathbf{f}^\star$ corresponding to the largest singular value.
\State Solve problem $\mathcal{P}_4$ for the phase rotation $\beta$ applied during projection (\ref{eq13}) to obtain the discrete phase solution $\widehat{\mathbf{f}}^\star$.
\Ensure Physically realizable analog Tx BF weights $\widehat{\mathbf{f}}^\star$.
\end{algorithmic}
\end{algorithm}

\begin{figure}[tp]
	\centering
	 \subfigure[Max SI power across Rx antennas.]
  {\includegraphics[height=0.18\textwidth] {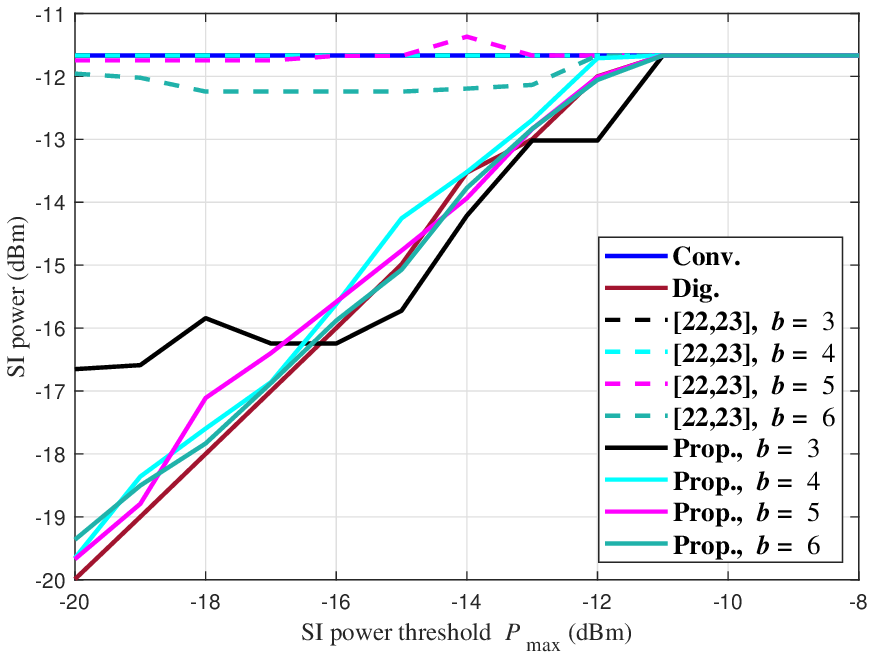}\label{SIC00}}\hfill
    \subfigure[Delivered Tx BF gain.]
    {\includegraphics[height=0.18\textwidth]{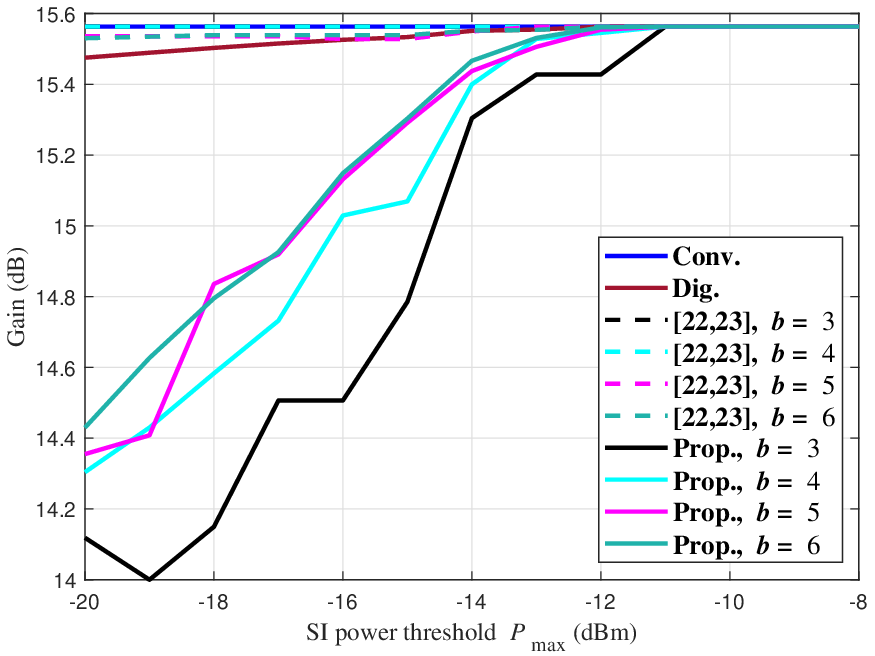}
	\label{gain00}}
        \caption{Performance of (a) maximum SI power across Rx antennas and (b) Tx BF gain of our proposed method versus benchmarks \cite{19,20} when the array is steered in the broadside direction.}

	\label{Prop}
\end{figure}

Now, we present numerical results to illustrate the advantages of our proposed method (denoted as Prop.). 
In Fig.~\ref{SIC00}, we evaluate per-antenna SIC performance by plotting the resulting maximum SI power across all 36 Rx antennas as a function of the SI power threshold $P_\mathrm{max}$. 
In Fig.~\ref{gain00}, we similarly plot the achieved Tx BF gain in the boresight direction $(\theta,\phi) = (0^\circ,0^\circ)$ as a function of $P_\mathrm{max}$. 
From Fig.~\ref{SIC00}, we can see that our proposed design can reliably cancel SI per-antenna to near or even below $P_\mathrm{max}$, whereas the benchmarks do not, which would subject Rx chain components to potential saturation. 
A commercial LNA \cite{lna}, for example, may only be linear for input powers up to about $-18$~dBm (its P1dB point). 
Using our proposed design with $P_\mathrm{max} \leq -19$~dBm could ensure SI power is kept sufficiently low to maintain linearity, assuming $b \geq 4$~bits. 
Under a strict threshold of $P_\mathrm{max} = -20$~dBm, for instance, our design guarantees SI power is kept about $8$~dB lower than \cite{19,20}, even with only $b=4$ bits of resolution.
Due to the coarse phase control with $b = 3$ bits of resolution, even with our proposed design, we cannot guarantee that SI power will be kept below $-17$~dBm; still, this is about $5$~dB lower than that with \cite{19,20} or CBF.

Fig.~\ref{gain00} illustrates that, while preserving the linearity of the Rx chain, our proposed design only sacrifices at most about $1.1$--$1.4$~dB in BF gain compared to the benchmarks. 
The reference scheme nearly approaches the maximum BF gain of $15.6$~dB achieved by CBF but, as seen in Fig.~\ref{SIC00}, this is at the cost of much higher SI. 
With our proposed approach, even strict thresholds of $P_\mathrm{max} = -20$~dBm can only result in about $1.1$~dB of loss in BF gain, and we can see that higher phase shifter resolution improves BF gain.
Our proposed approach sacrifices only $0.5$~dB of BF gain while satisfying a threshold of $P_\mathrm{max} = -16$~dBm, assuming $b \geq 4$ bits.
The superiority of our approach in its ability to maintain high gain while reliably reducing SI power per-antenna highlights the importance of explicitly accounting for discrete phase shifters during BF design, rather than making it an afterthought.
Naturally, since the fully-digital BF solution (Dig.) has more degrees-of-freedom, it can maintain both high gain and impressive SI reduction, but this is simply not attainable under analog BF with quantized phase shifters.

\begin{figure}[t]
	\centering
 \subfigure[Max SI power of proposed (dBm)]
  {\includegraphics[width=0.24\textwidth] {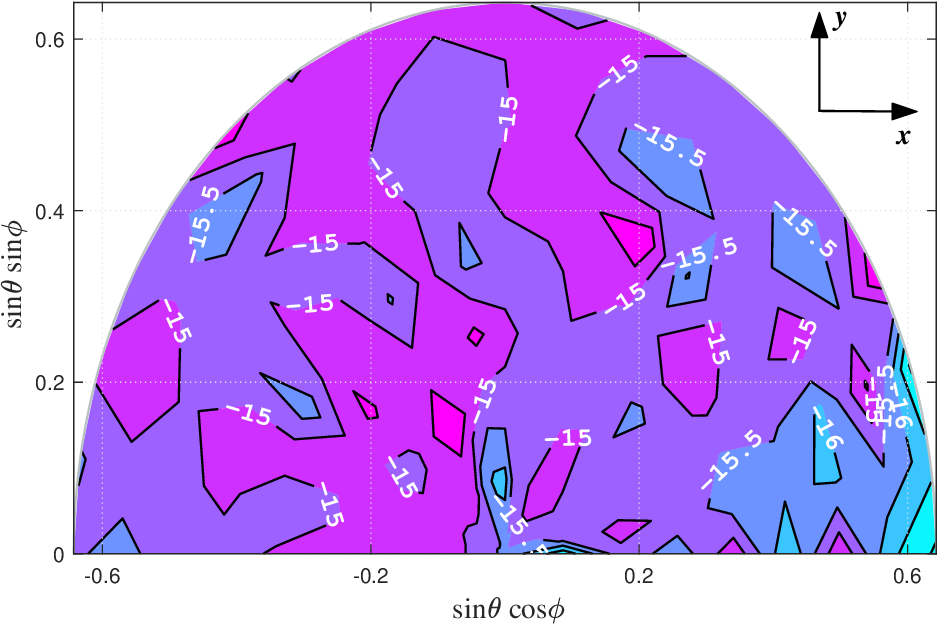}\label{SICprop}}\hfill
    \subfigure[Tx BF gain of proposed (dB)]
    {\includegraphics[width=0.24\textwidth]{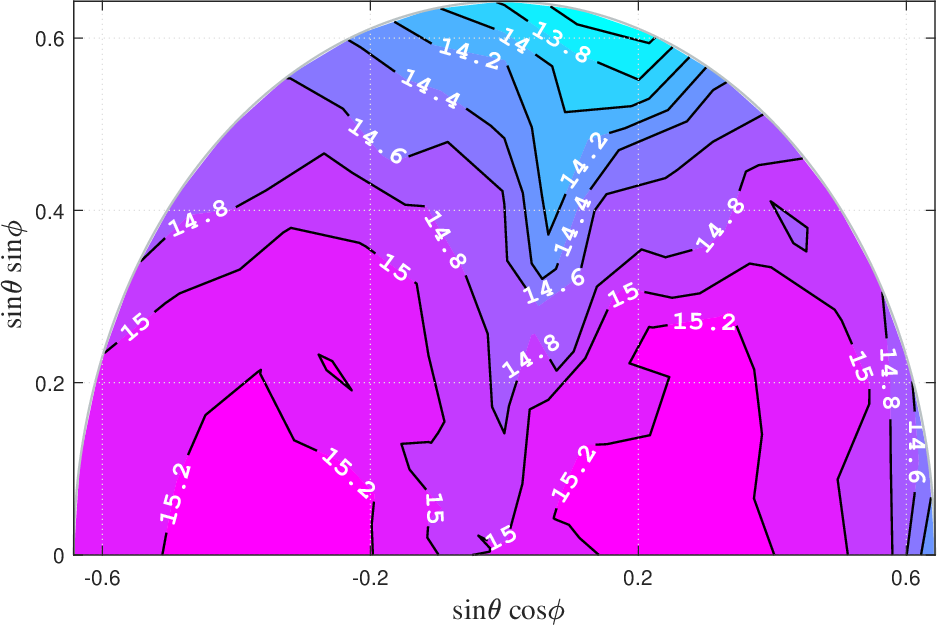}
	\label{gainprop}}
	 \subfigure[SI power difference (dB)]
  {\includegraphics[width=0.24\textwidth] {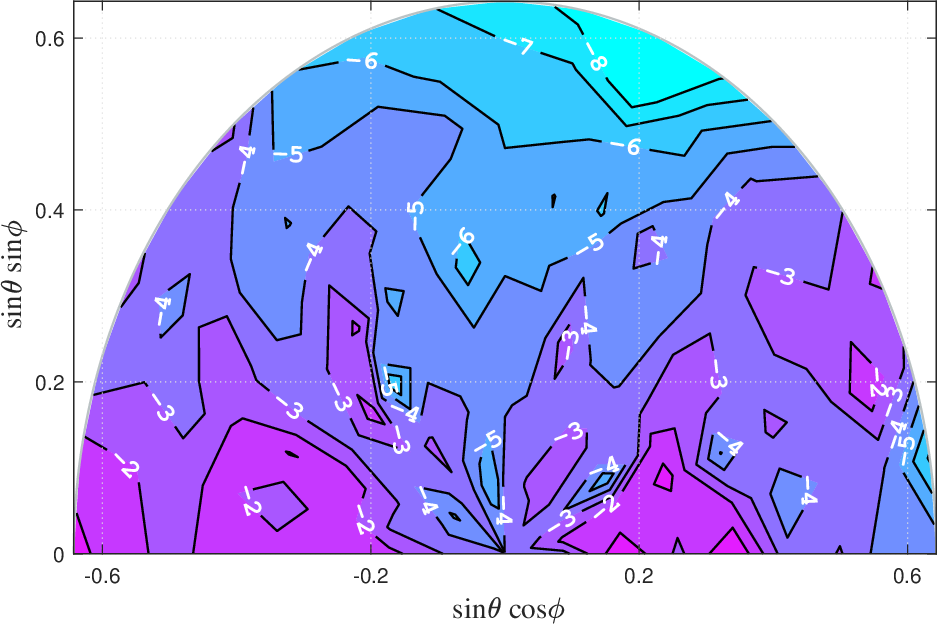}\label{SICdiff}}\hfill
    \subfigure[Tx BF gain difference (dB)]
    {\includegraphics[width=0.24\textwidth]{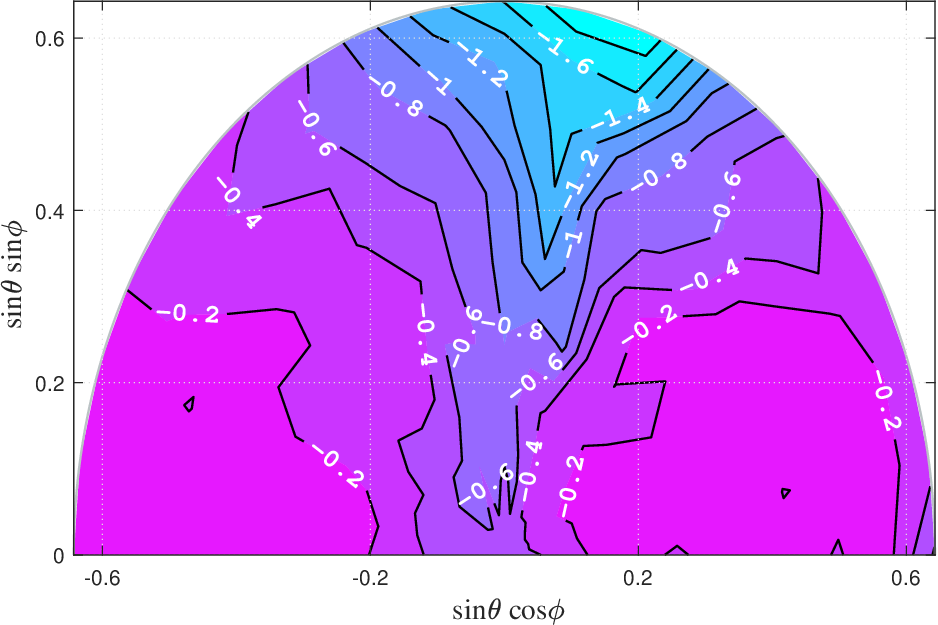}
	\label{gaindiff}}
        \caption{For various steering directions, the achieved (a) max SI power across Rx antennas and (b) Tx BF gain with the Prop. method, and differences in (c) max SI power and (d) Tx BF gain, relative to the reference method \cite{19,20}, where $b = 4$ bits and $P_\mathrm{max} = -16$~dBm.} 
	\label{Performance_diff}
\end{figure}

Finally, the contour plots of Fig.~\ref{Performance_diff} illustrate performance of our proposed method across a broad range of steering directions $(\theta,\phi)$ for $b=4$ bits and $P_\mathrm{max} = -16$ dBm. 
The results are demonstrated from the perspective of array surface, where positive direction of horizontal axes in Fig.~\ref{Performance_diff} denotes the direction of Tx subarray.
For conciseness, the top half of the steering directions are shown (i.e., $0^{\circ}\leq \phi \leq 180^{\circ}$) since the lower half is virtually symmetric. 
From Fig.~\ref{SICprop}, the Prop. method effectively suppresses SI power to near or below $P_\mathrm{max} = -16$~dBm for all considered directions, falling between $[-16,-15]$~dBm in most cases. 
In Fig.~\ref{gainprop}, we can see that the corresponding BF gain oscillates between $[14.6,15.35]$~dB for most considered directions, falling short by at most about $1$~dB compared to CBF.
We can see that steering upward (increasing $\theta$, $\phi \approx 90^{\circ}$) leads to slightly more loss in BF gain to maintain the per-antenna constraint and steering rightward away from Rx subarray ($\theta \approx 40^{\circ}$, $\phi \approx 0^{\circ}$) results in substantial SI reduction with negligible loss in BF gain. 
This is an artifact of the underlying structure of the SI channel and how it may dictate one's ability to both reduce SI and deliver high BF gain in certain directions. 

When compared to the benchmarks \cite{19,20} in Fig.~\ref{SICdiff} and Fig.~\ref{gaindiff}, our proposed method better reduces SI across all considered steering directions, with up to $8.4$~dB greater reduction.  
For most directions, our proposed design enjoys about $4$~dB lower SI power while only sacrificing about $0.4$~dB in Tx BF gain.
In particular, when steering rightward towards $\theta \approx 40^{\circ}, \phi \approx 0^{\circ}$, our proposed design enjoys up to about $7$~dB lower SI power while only sacrificing around $0.2$~dB in BF gain. 
On the contrary, when steering upward (increasing $\theta$), our proposed method enjoys low SI but sacrifices over $1.5$~dB of BF gain over the benchmarks \cite{19,20}.
Again, these are artifacts of the underlying structure of the SI channel $\mathbf{H}$ and highlight how exploring this further would be valuable future work.
In addition, these results illustrate that it may be preferable to adapt $P_\mathrm{max}$ based on steering direction to avoid prohibitive BF loss in less favorable directions.

\section{Conclusion}\label{Conclusion}
In this letter, we presented a novel Tx BF design for IBFD phased arrays which maximizes gain while ensuring SI power is kept below some tolerable threshold at each Rx antenna. 
Our design explicitly incorporates the non-convexity posed by limited resolution phase shifter components in analog BF systems. 
We accomplish this by transforming non-convex constraints into convex polygon constraints, which we solve using SDR and a rank refinement procedure.
Numerical results show that this approach much more reliably reduces SI while maintaining BF gain compared to traditional approaches which simply ignore the convexity during design and then perform a non-convex projection.
Valuable future work would explore how similar approaches may be employed in 5G/6G BF applications such as integrated sensing and communications.

\bibliography{IEEEabrv,refs.bib}

\end{document}